\documentclass[twoside]{article}

\usepackage{tabulary,graphicx,times,caption,fancyhdr,amsfonts,amssymb,amsbsy,latexsym,amsmath}
\usepackage[utf8]{inputenc}
\pdfoutput=1
\usepackage{url,multirow,morefloats,floatflt,cancel,tfrupee,textcomp,colortbl,xcolor,pifont}
\usepackage[nointegrals]{wasysym}
\usepackage{graphics}
\usepackage{ulem}
\usepackage{multirow}
\usepackage{float}
\usepackage{inputenc}
\usepackage{xspace}
\usepackage{url}
\usepackage{epstopdf}
\usepackage{amsfonts}
\usepackage{amsmath}
\usepackage{amssymb}
\usepackage{extarrows}
\graphicspath{ {./images/} }
\allowdisplaybreaks

\makeatletter

\usepackage{ifxetex}
\ifxetex\else
\usepackage{dblfloatfix}
\fi

\@ifundefined{subparagraph}{
\def\subparagraph{\@startsection{paragraph}{5}{2\parindent}{0ex plus 0.1ex minus 0.1ex}%
{0ex}{\normalfont\small\itshape}}%
}{}
\def\URL#1#2{\@ifundefined{href}{#2}{\href{#1}{#2}}}

\def\UrlOrds{\do\*\do\-\do\~\do\'\do\"\do\-}%
\g@addto@macro{\UrlBreaks}{\UrlOrds}

\makeatother

\usepackage[paperheight=11in,paperwidth=8.3in,margin=2.5cm,headsep=.7cm,top=2.5cm]{geometry}
\usepackage[T1]{fontenc}

\renewenvironment{abstract}
	{\trivlist\item[]\leftskip0pt\par\vskip4pt\noindent
  	\textbf{\abstractname}\mbox{\null}\\}
	{\par\noindent\endtrivlist}

\def\keywords#1{\par\medskip\par\noindent\textbf{Keywords}: #1\par}

 \date{} \emergencystretch 8pt

\makeatletter
\def\author#1{\gdef\@author{\hskip-\tabcolsep%
	\parbox{\textwidth}{\raggedright\bfseries#1\\[1pc]}}}
\def\address[#1]#2{\g@addto@macro\@author{\\\hskip-\tabcolsep\parbox{\textwidth}{\raggedright%
	\normalsize\normalfont\textsuperscript{#1}#2}}}
\let\addresslink\textsuperscript
\def\correspondence#1{\g@addto@macro\@author{\\\hskip-\tabcolsep\parbox{\textwidth}{\raggedright%
	\vspace*{10pt}\normalsize\normalfont~\\#1~\\[12pt]}}}
\def\email#1{\g@addto@macro\@author{\\\hskip-\tabcolsep\parbox{\textwidth}{\raggedright%
	\normalsize\normalfont Emails: #1}}}

\def\title#1{\gdef\@title{\vspace*{-30pt}%
	\raggedright\textbf{\@journaltitle}~\\%
  \raggedright\bfseries\ifx\@articleType\@empty\vspace*{20pt}\else%
  \vspace*{20pt}\@articleType\vspace*{20pt}\\\fi#1}}
\let\@journaltitle\@empty \def\journaltitle#1{\gdef\@journaltitle{{\normalfont\itshape#1}}}
\let\@articleType\@empty \def\articletype#1{\gdef\@articleType{{\normalfont\itshape#1}}}

\let\@runningHead\@empty \def\RunningHead#1{\gdef\@runningHead{{\normalfont #1}}}

\usepackage{fancyhdr}
\fancypagestyle{headings}{\fancyhf{}
  \fancyhead[R]{\itshape\@runningHead}
  \fancyfoot[C]{\thepage}}
\fancypagestyle{plain}{%
	\fancyhf{}\fancyhead[R]{Prepared for submission to Hindawi}
  \fancyfoot[C]{\thepage}}
\makeatother

\usepackage[numbers]{natbib}

\usepackage{float,xcolor}

\journaltitle{
Advances in High Energy Physics}
\articletype{Research Article} 

\begin{document}

\title{Strongly coupled fermions in odd dimensions and the running cut-off $\Lambda_d$}

\author{
		Evangelos G. Filothodoros\addresslink{1}}		
\address[1]{Institute of Theoretical Physics, Aristotle University of Thessaloniki, Thessaloniki, Greece.}

\correspondence{Correspondence should be addressed to 
    	Evangelos G. Filothodoros; efilotho@physics.auth.gr, vagfil79@gmail.com}



\maketitle 

\begin{abstract}
I study the fermionic $U(N)$ Gross-Neveu model at imaginary chemical potential and finite temperature for odd $d$ dimensions, in the strong coupling regime, by using the gap (saddle point) equation for the fermion condensate of the model. This equation describes the phase transitions from weak to strong coupling regime. I point out that the higher odd dimensional gap equations are linear combinations of the lower dimensional equations in a way that as the dimension of the model increases the lower dimensions are weaker coupled but still in the strong coupling regime. Interestingly, at a specific value of the chemical potential, exactly in the middle of the thermal windows that separate the fermionic from the bosonic (condensed) state of the fermions, I find the mass of the fermion condensate for $d=3,5,7,9$. An anomaly occurs at the $5$ dimensional theory where it is stronger coupled against other theories in higher dimensions and lower energy. The main idea of this work is that the cut-off $\Lambda$ regulator for the UV divergent parts of the fermion mass saddle point equation, plays the role of a physical parameter that makes the separation of the odd dimensional fermionic theories according to how deep they are in the strong coupling regime. This idea is based on the identity of the asymptotic freedom of the Gross-Neveu model as a toy model for QCD.

\keywords{Gross-Neveu; Strong coupling; Cut-off}
\end{abstract}

\section{Introduction}
One of the most extensively used model for the study of chiral symmetry, bosonisation physics \cite{Fradkin:1994tt} via  statistical transmutation \cite{Wilczek:1981du,Polyakov:1988md} and condensed matter physics, is the Gross-Neveu model and its generalizations in the presence of real and imaginary chemical potential. Recently it has been connected (see e.g. \cite{Karch:2016sxi,Murugan:2016zal,Seiberg:2016gmd,Kachru:2016rui,Meng:2020}) to particle-vortex duality e.g. \cite{Peskin:1977kp}. The extension of fermionic theories to finite temperature thermal field theory has also been considered (see  for example \cite{Giombi:2011kc,Aharony:2012ns} and references therein), in the context of various models that describe matter coupled to non-abelian Chern-Simons fields but also the more recent works about exploring the symmetry-breaking of conformal theories in the large-charge limit like \cite{Alvarez}. Interestingly, fermion condensation have been studied at higher dimensions \cite{Mello, David} as well as the anomalies \cite{Genolini, Gaiotto} in non-abelian gauge theories like the $5$ dimensional Yang Mills-theories, where we have instantons, associated with a global $U(1)$ symmetry. It seems that the Gross-Neveu model at imaginary chemical potential and finite temperature has an interesting identity at $5$ dimensions ($4+1$ dimensions for the compact dimension of temperature) that creates an anomaly of the original asymptotic freedom, arises from the gap equations of the model.

\section{Fermions coupled to Chern-Simons in a monopole background at $3d$}

When fermions are coupled to a Chern-Simons gauge field in a monopole background, unique characteristics arise within their system. These characteristics include the manifestation of anyonic statistics. These statistics differ from both fermionic and bosonic statistics, yet they can take fractional values. Exploring theories of this nature holds relevance in the realm of condensed matter physics, notably in phenomena like the Fractional Quantum Hall Effect (FQHE) and topological insulators. Instantons \cite{DiVecchia} become relevant during the analysis of an efficient field theory, such as the Chern-Simons theory, which elucidates the low-energy behavior of the Fractional Quantum Hall Effect (FQHE). They offer a structure for comprehending fractionalized excitations, representing intricate transitions between distinct topological sectors through nontrivial tunneling processes. Within these frameworks, one can investigate shifts between various fractional quantum Hall states, alterations in excitation statistics, and the creation of energy spectrum gaps.

In scenarios involving finite temperature, fermions coupled with a gauge field $B_\nu$ exhibit a temporal component $B_0$. This component akin to an imaginary chemical potential for the $U(1)$ charge, as elaborated in references such as \cite{ZinnJustin:2002ru}.  The work presented in \cite{Filothodoros:2016txa} proposed that in the context of three dimensional Eucliden space, Dirac fermions coupled to an abelian Chern-Simons field at level $k$, exhibit a significant link between the existence of a monopole charge and the existence of an imaginary chemical potential \footnote{My notations follow \cite{ZinnJustin:2002ru}.}. This relationship can be comprehended by examining the subsequent fermionic partition function at finite temperature
\begin{align}
\label{DiracCSPF}
Z_{fer}(\beta,k)&=\int [{\cal D}B_\nu][{\cal D}\bar\psi][{\cal D}\psi]\exp{\left[-S_{fer}(\bar\psi,\psi,B_\nu)\right]}\,,\\
\label{Sf}
S_{fer}(\bar\psi,\psi,B_\nu)&=-\int_0^{\beta}\!\!\!dx^0\!\!\int \!\!d^2\bar{x}\left[\bar{\psi}(\slash\!\!\!\partial -i\slash\!\!\!\!B)\psi+i\frac{k}{4\pi}\epsilon_{\nu\lambda\rho}B_\nu\partial_\lambda B_\rho+...\right]\,.
\end{align}
Additionally, there exist fermionic self-interactions denoted by dots. I proceed by expanding the Chern-Simons field around a static monopole configuration, which is independent of time $\bar{B}_\nu$ \cite{Fosco:1998cq}
\begin{equation}
\label{Aexpansion}
B_\nu=\bar{B}_\nu+ b_\nu\,,\,\,\,\bar{B}_\nu=(0,\bar{B}_1(\bar{x}),\bar{B}_2(\bar{x}))\,,\,\,\, b_\nu=(b_0(x^0),b_1(x^0,\bar{x}),b_2(x^0,\bar{x}))\,,
\end{equation}
which is normalized as\footnote{For example, one may consider the theory on $S^1\times S^2$.}

\begin{equation}
\label{monopole}
\frac{1}{2\pi}\int d^2x \bar{F}_{12}=1\,,\,\,\,\bar{F}_{\nu\lambda}=\partial_\nu \bar{B}_\lambda-\partial_\lambda \bar{B}_\nu\,
\end{equation}
and $b_\nu$ is a backround gauge field.
Therefore, equation (\ref{Sf}) encapsulates the potential for monopole configurations within the fermionic theory. This corresponds to the scenario where fermions are associated with the attachment of $k$ units of monopole charge as
\begin{equation}
\label{Sfexp}
S_{fer}(\bar\psi,\psi,B_\nu) =-\int_0^\beta \!\!\!dx^0\!\!\int \!\!d^2\bar{x}\left[\bar\psi(\slash\!\!\!\partial-i\gamma_i\bar{B}_i-i\gamma_\nu b_\nu)\psi+i\frac{k}{4\pi}\epsilon_{\nu\lambda\rho}b_\nu\partial_\lambda b_{\rho}+..\right]-ik\int_0^\beta \!\!\!dx^0 b_0\,.
\end{equation}
We can carry out the path integral over the Chern-Simons fluctuations by focusing on a sector characterized by a fixed total monopole charge. To achieve this, I assume the existence of a mean field approximation within this sector, where the spatial fluctuations of the Chern-Simons field balance the magnetic background gauge field, represented as $\langle b_i\rangle =-\bar{B}_i$ \cite{Barkeshli:2014ida}. \footnote{ Further clarification might be valuable regarding the potential requirement of an appropriate large-$N$ for the validity of this approximation.} This is akin to envisaging a reduction in which the integral of the background gauge field along the thermal circle remains constant. Subsequently, I derive the following expression:
\begin{align}
Z_{fer}(\beta,k)&=\int [{\cal D} b_0][{\cal D}\bar\psi][{\cal D}\psi]\exp{\left[\int_0^\beta \!\!\!dx^0 \!\!\int \!\!d^2\bar{x}\left[\bar\psi(\slash\!\!\!\partial-i\gamma_0 b_0)\psi+..\right]+ik\int_0^\beta \!\!\!dx^0 b_0\right]}\nonumber \\
\label{DiracCSPFfin}
&=\int ({\cal D}\theta)e^{ik\theta}Z_{gc,fer}(\beta,-i\theta/\beta)\,,
\end{align}
where  $\theta=\int_0^\beta dx^0 b_0(x^0)$, $Z_{gc,fer}(\beta,-i\theta/\beta)$ is the grand canonical partition function for the fermionic theory and I have used standard formulae from \cite{ZinnJustin:2002ru}. We see that the CS level $k$ plays the role of the eigenvalue $Q$ of the $U(1)$ charge operator. 
An essential inference can be drawn: the partition function at finite temperature for Dirac fermions coupled to an abelian Chern-Simons gauge field with a level $k$ in a monopole background is tantamount to the canonical partition function of the fermions with a fixed fermion number of $k$.

The preceding discourse underscores the close interconnection between the partition function of fermions coupled to an abelian Chern-Simons field in a monopole environment and the corresponding canonical partition function characterized by a constant total $U(1)$ charge. This charge, identified as the instanton number, is borne by instanton configurations within the theory, akin to particles in a $5d$ framework. This analogy resonates with the earlier illustration of monopole operators in $3d$ gauge theories \cite{Genolini}. 

The $5d$ theories, resembling a Yang-Mills theory with a background gauge field, exhibit infrared freedom and feature a singularity pole in the ultraviolet regime, necessitating the implementation of a cut-off regulator denoted as $\Lambda$. In this study, our focus lies in delving into the conceptual significance of this cut-off, accomplished through an exploration of the gap equations pertaining to the fermionic Gross-Neveu model operating under an imaginary chemical potential and at finite temperature (or through the process of dimensional reduction by considering the theory on a circle).

\section{The $U(N)$ fermionic Gross-Neveu at imaginary chemical potential and finite temperature in odd dimensions - The gap equations}

Within this study, I will analyze a fermionic theory characterized by an odd dimension $d$, such as the Gross-Neveu model, while considering an imaginary chemical potential (linked to the temporal gauge field) and finite temperature. Concurrently, I will explore the manifestation of an anomaly within the $5d$ theory. Notably, the Gross-Neveu model serves as a simplified representation of Quantum Chromodynamics (QCD) and its property of asymptotic freedom. It is well-established that the fermionic Gross-Neveu model displays distinct patterns of symmetry breakdown at finite temperature $T$, especially in the absence of a chemical potential $U(N)$. At lower temperatures, the model enters a phase characterized by broken parity symmetry. This phase diminishes as the temperature approaches a critical value.
Nonetheless, my prior work in \cite{Filothodoros:2016txa} demonstrated a shift in this scenario owing to the presence of an imaginary chemical potential and its consequent influence on the model's phase structure. It was further observed in the same study that the renowned Bloch-Wigner function \cite{Zagier1} bears relevance in our calculations of the gap equations and free energies. In the case of odd dimensions $d$, these calculations can be expressed as finite sums of Nielsen's generalized polylogarithms \cite{Filothodoros:2016txa, Borwein, Kolbig}. However, the expressions become significantly more intricate for even dimensions $d$. Moreover, I have previously noted the connection between $1d$ theories and the physics of the $3d$ model, as well as the connection of $1d$ and $3d$ theories to the physics of the $5d$ model, and so forth. Given these reasons, my focus will be directed towards exploring the extensions of the $3d$ model into higher odd dimensions ($d>3$). The gap equations for the model in arbitrary odd dimensions can be articulated as follows:

\subsection{The gap equations}
The Gross-Neveu (GN) model in a Euclidean space with $d$ dimensions is characterized by the extension of the $3d$ action \cite{Filothodoros:2018, Thesis Filothodoros, Petkou:1998wd, Christiansen:1999uv}. In this context, I perform a dimensional reduction of the $d$-dimensional theory over the thermal circle, yielding an effective theory in $d-1$ dimensions. This effective theory remains applicable for distances significantly surpassing the radius $\beta=\frac{1}{T}$, where $\beta$ is the inverse temperature.

\begin{align}
 S_{fer} = -\int_0^\beta \!\!\!dx^0\int \!\!d^{d-1}\bar{x} \left[\bar{\psi }^{i}(\slash\!\!\!\partial  -i\gamma_0 b)\psi ^{i}
+\frac{G_d}{2({\rm Tr}\mathbb I_{d-1})N}\left (\bar{\psi }^{i}\psi ^{i}\right )^{2} +ib NQ_d\right]\,,
\end{align}
with  $Q_d$ the $N$-normalized $d$-dimensional fermionic number density and $i=1,2,..N$ that plays the role of the eigenvalue of the $U(1)$ charge operator. For odd $d$ we take the dimension of the gamma matrices to be ${\rm Tr}{\mathbb I}_{d-1}=2^{\frac{d-1}{2}}$.

The $d$-dimensional  gap equation for the fermionic condensate becomes (by introducing the auxiliary scalar fields $m_*$ and $b_*$):
\begin{align}
\frac{m_*}{G_d}&=\frac{m_*}{\beta}\sum_{n=-\infty}^\infty\int^\Lambda\!\!\frac{d^{d-1} \bar{p}}{(2\pi)^{d-1}}\frac{1}{\bar{p}^2+(\omega_n-b_*)^2+m_*^2}\,
\end{align}

The primary challenge encountered when considering the Gross-Neveu (GN) model in dimensions greater than $3$ ($d>3$) is that the gap equation acquires a finite count of higher-order divergent terms as the ultraviolet cut-off $\Lambda$ approaches infinity. Unlike the scenario in $d=3$, it's not feasible to solely manage these divergences through the adjustment or renormalization of a single coupling constant $G_d$. Consequently, we must address this particular matter by employing an alternative renormalization technique.

In the subsequent discussion, I will delve into specific cases to provide a comprehensive understanding of the higher-dimensional models. I will elaborate on the instances where $d=5$, $d=7$, and $d=9$ to highlight some of the overarching traits of these models. Beginning with the case of $d=5$ and building upon the findings presented in references such as \cite{Filothodoros:2016txa, Filothodoros:2018}, we can outline the structure of the gap equation as follows:
\begin{align}
\label{ge5}
m_*\left[-({\cal M}_5\beta^3+D_3(-z_*))-
\frac{1}{2!}\ln^2\!|z_*|\left(D_1(-z_*)-\frac{2\Gamma}{3\pi}\right)\right]&=0\,
\end{align}

where 
\begin{equation}
\frac{{\cal M}_5}{(2\pi)^2}=\frac{1}{G_{5,*}}-\frac{1}{G_5}\,,\,\,\,\,\Gamma=\Lambda\beta\,
\end{equation}
(for the cut-off $\Lambda$ regulator I will examine a new approach in the next subsection) and $z_*=e^{-ib_*\beta-m_*\beta}$.

In the context of the equation (\ref{ge5}), $G_5$ represents the bare coupling, while $G_{5,*}$ signifies the critical coupling at zero temperature. Should ${\cal M}_5$ possess a value greater than zero, it implies that $G_5$ exceeds $G_{5,*}$, thereby entering the domain of strong coupling. Consequently, the existence of a non-zero solution for $m_*$ in equation (\ref{ge5}) would lead to the breaking of parity symmetry by giving mass to elementary fermions.

In deriving the gap equation, I have omitted an infinite number of terms that scale inversely with powers of $\Lambda$, and I have also observed that the last term within the parentheses mirrors the analogous gap equation in three dimensions. Notably, we discern indications of a partial deconstruction of the higher-dimensional models, where they can be expressed in terms of lower-dimensional quantities.

As we move into dimensions $d>3$, a pivotal aspect emerges in the explicit inclusion of the cut-off within the gap equation. This becomes evident when comparing the $5d$ case with the $3d$ counterpart. I have underscored the fact that ${\cal M}_5$ remains independent of the cut-off $\Lambda$. As a result, for a given temperature, the $5d$ gap equation transforms into a two-parameter equation for $z*$. This implies that there is no clear-cut method to manipulate the single coupling constant of the theory, represented by the parameter ${\cal M}_5$, in a manner that yields a result free from cut-off dependence. This is indicative of the nonrenormalizability of the 5-dimensional theory unless an alternate method is employed.

When introducing an imaginary chemical potential, the situation grew more intriguing. I encountered once again nontrivial roots of $D_3(-z_*)$ situated on the unit circle. This discovery enabled me to explore the critical theory while setting ${\cal M}_5$ to zero. Through a brief exploration using Mathematica, I identified two zeros of $D_3(-z)$ on the unit circle. These zeros hold significance in determining the thermal windows where the system undergoes bosonization. Interestingly, in \cite{Filothodoros:2018}, I approximated their positions to a high degree of precision, and their values were found to be rational multiples of $\pi$, as highlighted in a comprehensive study \cite{Etienne}, as:
\begin{align}
\nonumber D_3(-e^{-i\beta b_*})=Cl_3(\beta b_*\pm\pi)=0\Rightarrow\\ \beta b_*\approx \frac{7\pi}{13}\,{\rm or}\,\frac{19\pi}{13} \,\,\,({\rm mod}\,2\pi)\,.
\end{align}
Using the periodic properties of the Clausen functions, the relevant results are 
\begin{equation}
Cl_3\left(\frac{6\pi}{13}\right)=Cl_3\left(\frac{20\pi}{13}\right)=0.000362159\,.
\end{equation}

I observed that this pattern extends across all odd dimensions. The charge becomes zero, and the system undergoes bosonization. Nevertheless, in distinction from the analogous scenario in $d=3$, a non-zero solution for $m_*$ in the critical condition ${\cal M}_5=0$ becomes contingent on the arbitrary parameter $\Gamma$.

Recalling the seven-dimensional case, which effectively demonstrates how my findings hold true for higher dimensions \cite{Filothodoros:2018}, the gap equation can be expressed as follows:
\begin{align}
m_*\left[(-{\cal M}_7\beta^5+D_5(-z_*))+\frac{1}{3!}\ln^2\!|z_*|\left(D_3(-z_*)+\frac{\Gamma^3}{45\pi}\right)
+\frac{1}{4!}\ln^4\!|z_*|\left(D_1(-z_*)-\frac{4\Gamma}{15\pi}\right)\right]=0\,
\end{align}
where the parameter $\Gamma$ has been defined above, 
and 
\begin{equation}
\frac{3{\cal M}_7}{(2\pi)^3}=\frac{1}{G_{7,*}}-\frac{1}{G_7}\,.
\end{equation}

As in previous case, it's evident from the initial equation of the $d=7$ scenario that terms corresponding to the respective three- and five-dimensional gap equations are present.

Additionally, as previously mentioned, we observe that attempting to adjust $\Gamma$ to eliminate the constant terms $D_3(-1)$ and $D_1(-1)$ in the expansion of the gap equation near $m_*=0$, and consequently achieve a definite multicritical behavior for the effective action, turns out to be unfeasible. This challenge persists clearly for all dimensions $d>7$.

Proceeding to the case of a non-zero chemical potential, we can seek the roots of the critical gap equation located on the unit circle. Once again, their positions are strikingly well approximated, even more precisely than in the $d=5$ case, by rational multiples of $\pi$, as follows:
\begin{align}
\nonumber D_5(e^{-i\beta b_*})=Cl_5(\beta b_*\pm \pi)=0\Rightarrow \\
 \beta b_*\approx \frac{26\pi}{51}\,{\rm or}\,\frac{76\pi}{51} \,\,\,({\rm mod}\,2\pi)\,.
\end{align}
The relevant result is 
\begin{equation}
Cl_5\left(\frac{25\pi}{51}\right)=Cl_5\left(\frac{77\pi}{51}\right)=0.000129657\,
\end{equation}

The $9d$ gap equation in the same way is:

\begin{align}
m_*[-({\cal M}_9\beta^7+D_7(-z_*))-\frac{\ln^2|z_*|}{4!}(D_5(-z_*)-\frac{4\Gamma_1^5}{525\pi})-\\
-\frac{\ln^4|z_*|}{5!}(D_3(-z_*)+\frac{4\Gamma_2^3}{315\pi})-\frac{\ln^6|z_*|}{6!}(D_1(-z_*)-\frac{16\Gamma_3}{175\pi})]=0
\end{align}

where $\Gamma=\Lambda\beta$ the cut-off and

\begin{equation}
\frac{15{\cal M}_9}{(2\pi)^4}=\frac{1}{G_{9,*}}-\frac{1}{G_9}.
\end{equation}

An intriguing observation, as highlighted in the Introduction, is that the gap equation for higher odd dimensions (such as the $9d$ case) can be expressed as a linear combination of the equations from lower dimensions ($3d$, $5d$, and $7d$). This relationship can be articulated as follows:

\begin{equation}
m_*(-g_9-\frac{\ln^2|z_*|}{4!}g_7
-\frac{\ln^4|z_*|}{5!}g_5-\frac{\ln^6|z_*|}{6!}g_3)=0
\end{equation}
and for $m_*\not=0$
\begin{equation}
\label{gaplinear9}
g_9=-\frac{\ln^2|z_*|}{4!}g_7
-\frac{\ln^4|z_*|}{5!}g_5-\frac{\ln^6|z_*|}{6!}g_3
\end{equation}
where $g_d$ the gap equation for $d=3,5,7,9$. I believe that this result could also be repeated in the case of fermions coupled with a non-Abelian Chern-Simons field in a suitable gauge approximation, constituting a generalization of \cite{Aharony:2012ns} for $d=5,7,9,...$.

Again for non-zero values of the chemical potential we have the thermal window's opening and closing values of $\beta b_*$

\begin{align}
\nonumber D_7(e^{-i\beta b_*})=Cl_7(\beta b_*\pm \pi)=0\Rightarrow \\
 \beta b_*\approx \frac{103\pi}{205}\,{\rm or}\,\frac{307\pi}{205} \,\,\,({\rm mod}\,2\pi)\,.
\end{align}
The relevant result is 
\begin{equation}
Cl_7\left(\frac{102\pi}{205}\right)=Cl_7\left(\frac{308\pi}{205}\right)=-0.000101475\,
\end{equation}

The fundamental characteristics outlined earlier remain consistent as we transition to higher dimensions. The partial deconstruction of $d$-dimensional gap equations into components from lower dimensions persists. In specific terms, the $d$-dimensional gap equation incorporates elements from gap equations of dimensions $d-2, d-4,...,5,3$.

The thermal windows are:\\
	
	{\centering
		\begin{tabular}{|c|c|c|}
			\hline 
			\multicolumn{3}{|c|}{\textbf{Table 1. Thermal Windows for the Gross-Neveu model}} \\ 
			\hline 
			Dimensions & Closing T & Opening T \\ 
                        \hline
                        3 & $\frac{3b_*}{4\pi}$ & $\frac{3b_*}{2\pi}$\\
			\hline 
			5 & $\frac{13b_*}{19\pi}$ & $\frac{13b_*}{7\pi}$ \\ 
			\hline 
			7 & $\frac{51b_*}{76\pi}$ & $\frac{51b_*}{26\pi}$ \\ 
			\hline 
			9 & $\frac{205b_*}{307\pi}$ & $\frac{205b_*}{103\pi}$ \\ 
			\hline 
			11 & $\frac{819b_*}{1228\pi}$ & $\frac{819b_*}{410\pi}$ \\ 
			\hline 
		\end{tabular}\par} 

Let's direct our attention to the scenario concerning the thermal window of the $3d$ theory. The borders for $b_*$ are $2\pi T/3$ and $4\pi T/3$. These are the points where $D_2(-z*)$ \cite{Zagier1, Zagier2} takes its maximum value (imaginary part) on the unit circle. On the unit circle $D_2(-z*)=Cl_2(\pi\pm\beta b_*)$. At the middle point of the thermal windows $D_{even}(-z*)$ is zero and we are at the full bosonisation points.

\subsection{The running cut-off $\Lambda_d$}

The concept of a running cut-off $\Lambda_d$ in the context of quantum field theory refers to the idea of using a momentum (energy) dependent scale in the regularization of divergent loop integrals. Unlike a fixed cut-off $\Lambda$, which imposes a universal upper limit on momenta in all processes, a running cut-off varies with the momentum involved in a specific process \cite{Litim, Chu, Epelbaum, Barlas}. The purpose of using a running cut-off is to better capture the energy scale of the physics being probed in a given calculation.

In quantum field theory, loop diagrams involving virtual particles can lead to divergent integrals that do not produce meaningful results without regularization. A fixed cut-off involves truncating momenta above a fixed scale $\Lambda$, but this can be overly simplistic, as different processes might involve vastly different energy scales.

By using a running cut-off $\Lambda_d$, the regularization takes into account the fact that different loop momenta contribute differently to the overall process depending on their energy scale. At low momenta, the cut-off is effectively larger, allowing for a broader range of momenta to contribute. At high momenta, the cut-off is smaller, regulating the potentially divergent behavior of the loop integrals.

The running cut-off may be used in theoretical studies, especially in the context of effective field theories or models that aim to describe physics at different energy scales, like the Gross-Neveu model at finite temperature and for arbitrary odd dimensions of this work. However, it introduces its own complexities and challenges, particularly in the context of renormalization, where the regulator's behavior must be carefully controlled to ensure that physical predictions remain consistent and meaningful.

The running cut-off concept is particularly relevant when trying to bridge the gap between high-energy and low-energy descriptions of particle interactions and when dealing with strongly interacting systems like Quantum Chromodynamics (QCD). We have to explore different functional forms for the running cut-off in order to achieve physically meaningful and consistent results in our calculations. In the next section I will proceed to a completely new method of regularization for theories with fermions when $d>3$.

\section{Strongly coupled fermions and anomaly of the $5d$ gap equation}
 To calculate the condensate gap-equation of the $U(N)$ Gross-Neveu model in arbitrary odd dimensions and in the presence of imaginary chemical potential $\mu=-ib$ I use the Euclidean action from \cite{Filothodoros:2018}.
 \begin{equation}
 	S_{fer} = -\int_0^\beta \!\!\!dx^0\int \!\!d^{d-1}\bar{x} \left[\bar{\psi }^{i}(\slash\!\!\!\partial  -i\gamma_0 b)\psi ^{i}+\frac{G_d}{2({\rm Tr}\mathbb I_{d-1})N}\left (\bar{\psi }^{i}\psi ^{i}\right )^{2} +ib NQ_d\right]\,,\,\,\,i=1,2,..N.
 \end{equation}
where $Q_d$ is the eigenvalue of the $N$-normalized fermion number density operator $\hat{Q}_d=\psi^{i\dagger}\psi^i/N$ in $d$ odd dimensions and it comes from the Lagrangian of the model that possesses a $U(1)$ global symmetry. Introducing an auxiliary scalar field $m$ the canonical partition function is given by
 \begin{equation}
 	S_{fer,eff}=iQ_d\int_0^\beta \!\!\!dx^0\!\!\int\!\!d^{d-1}\bar{x}\,b-\frac{{\rm Tr}\mathbb I_{d-1}}{2G_d}\int_0^\beta \!\!\!dx^0\int d^{d-1}\bar{x} m^2+\rm Tr\ln(\slash\!\!\!\partial-i\gamma_0b+m)_\beta
 \end{equation}
 
 To evaluate the condensate gap equation I look for constant saddle points $b_*$ and $m_*$. At large-$N$ we have the gap equation
 \begin{equation}
 	\frac{\partial}{\partial m}S_{fer,eff}\Biggl|_{m_*,b_*}\!\!\!\!=0\,\,\,\Rightarrow\,\,\, -\frac{m_*}{G_d}+\frac{m_*}{\beta}\sum_{n=-\infty}^\infty\int^\Lambda\!\!\frac{d^{d-1} \bar{p}}{(2\pi)^{d-1}}\frac{1}{\bar{p}^2+(\omega_n-b_*)^2+m_*^2}=0
 \end{equation}
 where  the fermionic Matsubara sums are over the discrete frequencies $\omega_n=(2n+1)\pi/\beta$. The divergent integrals are regulated by the cut-off $\Lambda$. 
 The main issue with the GN model in $d>3$ is that the gap equation has  a finite number of higher order divergent terms as $\Lambda\rightarrow\infty$, which cannot be simply taken care of by the adjustment/renormalization of the single coupling $G_d$. In order to calculate the above integral of the general $d$ gap equation I use a standard inversion formula for the hypergeometric function which allows me to obtain the cut-off $\Lambda$ part 
\begin{align}
\label{Cutoff}
\int^\Lambda\!\!\!\frac{d^d p}{(2\pi)^d}\frac{1}{p^2+m*^2} &= \frac{1}{G_{d,*}}-\frac{\Lambda^{d-2}}{d-2}\frac{S_d}{(2\pi)^d}{}_2F_1\left(1,\frac{d}{2}-1;\frac{d}{2};-\frac{\Lambda^2}{m*^2}\right)\nonumber\\
&\hspace{-2cm}=\frac{1}{G_{d,*}}-\frac{S_d}{(2\pi)^d}\left[\Gamma\left(\frac{d}{2}\right)\Gamma\left(2-\frac{d}{2}\right)\frac{m*^{d-2}}{d-2}+\frac{m^2_*\Lambda^{d-4}}{d-4}{}_2F_1\left(1,2-\frac{d}{2};3-\frac{d}{2};-\frac{m*^2}{\Lambda^2}\right)\right]
\end{align}
where $S_d=2\pi^{d/2}/\Gamma(d/2)$. This way we  see that for odd $d$ there are a finite number of divergent terms as $\Lambda\rightarrow\infty$. To avoid these divergent parts I have to try a kind of regularization (similar to \cite{Filothodoros Lattices}) where the gap equation is $D_{odd}(-z*)=0$ at the highest odd dimension and at the critical point of the corresponding field theory. The lower dimensions have their own gap equations inside the gap equation of the $d$ model. A new approach gives to the cut-off $\Lambda$ different values depending on the dimension of the gap equation. It is like we change the bare cut-off $\Lambda$ with a running cut-off $\Lambda_d$ that varies as a function of energy (see for example \cite{ Epelbaum, Barlas, Tao}). It takes into account the renormalization group flow of the theory, which describes how the theory's parameters evolve as the energy scale changes, instead of the bare cut-off which is a fixed regularization scale introduced at the beginning of a calculation to handle UV divergences. Let's see for example ({\ref{Cutoff}}) at $d=7$. The last hypergeometric term gives us the sum:

\begin{equation}
\label{Parts}
\frac{m_*^2\Lambda_d^3}{3}-m_*^4\Lambda_d-\frac{m_*^6}{\Lambda_d}
\end{equation}

For large energies (large running cut-off) the last term of ({\ref{Parts}}) goes to zero so, we may always regulate the theory of higher dimension in higher energy in order to eliminate the $\frac{1}{\Lambda_d}$ part. On the other hand the other terms are infrared free, so we have to regularize the theory. As the dimension decreases the running cut-off (energy) increases and the theory is at its weak coupling. On the other hand as the dimension increases the running cut-off decreases and the fermions are more stronger coupled according to asymptotic freedom. Similar renormalisation methods have been used in \cite{Evans, Grozdanov} with a hard momentum cut-off with $p^2<\Lambda^2$, where the only degrees of freedom are those with momentum less than the cut-off. These methods have application in 
determining the correspondence between the Wilsonian cut-off scale on the boundary and its holographically dual bulk theory.
The critical gap equation $({\cal M}_d=0)$ in odd dimensions that determines the condensate mass is (I assume that we are at $T=0$ where we substitute the bare coupling with its renormalized version):

\begin{equation}
-{\cal M}_d\beta^{d-2}+D_{d-2}(-z_*)=0\rightarrow D_{d-2}(-z_*)=0
\end{equation}
for $d=3,7,11,...$

and 

\begin{equation}
{\cal M}_d\beta^{d-2}+D_{d-2}(-z_*)=0\rightarrow D_{d-2}(-z_*)=0
\end{equation}
for $d=5,9,13,...$. 
 
 \begin{itemize}
 	\item \underline{3d gap equation}
 \end{itemize}
 
 The gap equation in 3 dimensions of the Gross-Neveu model at imaginary chemical potential at the critical point ${\cal M}_3=0$ or $G_3=G_{3,*}$ and for $b_*=\pi/\beta$, $m_*\not=0$ turns to:
 
 $m_*\left[-{\cal M}_3\beta+D_1(e^{-m_*\beta})\right]=0\,\rightarrow m_* D_1(e^{-m_*\beta})=0\rightarrow D_1(e^{-m_*\beta})=0\rightarrow \ln(1-e^{-m_*\beta})-\ln\frac{|e^{-m_*\beta}|}{2}=0$. Then: 
 \begin{equation}
 \boxed
 {m_*\beta=\ln\phi^2}
 \end{equation}
 \cite {Filothodoros:2016txa,Sachdev:1993pr}.
  Also $-z_*=e^{-m_*\beta}$ at $b_*=\pi/\beta$ and $\beta=\frac{1}{T}$.

 \begin{itemize}
 	\item \underline{5d gap equation}
 \end{itemize}
 
 The gap equation in 5 dimensions of the Gross-Neveu model at imaginary chemical potential for $b_*=\pi/\beta$, $m_*\not=0$ turns to:
 
\begin{equation}
m_*[-({\cal M}_5\beta^3+D_3(e^{-m_*\beta}))-\frac{\ln^2|z_*|}{2!}(D_1(e^{-m_*\beta})-\frac{2\Gamma}{3\pi})]=0
\end{equation}
where $\Gamma=\Lambda\beta$ the cut-off. Obviously the gap equation includes the $D_3$ part for the $5d$ theory and the $3d$ gap equation with the $D_1$ part. If we set $\Gamma=\Lambda_3\beta$, put the $5d$ theory at the critical point ${\cal M}_5=0$ or $G_5=G_{5,*}$ and assume that $D_1(e^{-m_*\beta})-\frac{2\Lambda_3\beta}{3\pi}=0$, we find from gap equation $D_3(e^{-m_*\beta})=0$: 
\begin{equation}
\boxed
{m_*\beta=2.03185}
\end{equation}
and from the $3d$ equation that $\Lambda_3\beta=4.12525$. 
 
We see that $D_1(e^{-m_*\beta})-\frac{2\Lambda_3\beta}{3\pi}=0$ is the $3d$ gap equation so we may change $\frac{2\Lambda_3\beta}{3\pi}$ with ${\cal M}_3^{(5)}\beta$ which is the mass scale that separates weak from strong coupling of the $3d$ theory included in the $5d$ theory. 
 
 \begin{itemize}
 	\item \underline{7d gap equation}
 \end{itemize}
 
 The gap equation in 7 dimensions of the Gross-Neveu model at imaginary chemical potential and for $b_*=\pi/\beta$, $m_*\not=0$ turns to:
 
\begin{equation}
m_*[(-{\cal M}_7\beta^5+D_5(e^{-m_*\beta}))+\frac{\ln^2|z_*|}{3!}(D_3(e^{-m_*\beta})+\frac{\Gamma_1^3}{45\pi})+\frac{\ln^4|z_*|}{4!}(D_1(e^{-m_*\beta})-\frac{4\Gamma_2}{15\pi})]=0
\end{equation}
where $\Gamma=\Lambda\beta$ the cut-off. Again the gap equation includes the $D_5$ part for the $7d$ theory and the $5d$ and $3d$ equations with $D_3$ and $D_1$ parts respectively. If we set $\Gamma_2=\Lambda_3\beta$ and $\Gamma_1=\Lambda_5\beta$ and put the $7d$ theory at the critical point ${\cal M}_7=0$ or $G_7=G_{7,*}$, we find from gap equation $D_5(e^{-m_*\beta})=0$:
\begin{equation}
\boxed 
{m_*\beta=2.89218}
\end{equation}
and that $\Lambda_5\beta=6.0403$ from $D_3(e^{-m_*\beta})+\frac{(\Lambda_5\beta)^3}{45\pi}=0$ and $\Lambda_3\beta=16.3642$ from $D_1(e^{-m_*\beta})-\frac{4\Lambda_3\beta}{15\pi}=0$.

We will also change (like the $5d$ case) $\frac{4\Lambda_3\beta}{15\pi}$ with ${\cal M}_3^{(7)}\beta$ and $\frac{(\Lambda_5\beta)^3}{45\pi}$ with ${\cal M}_5^{(7)}\beta^3$, which are the mass scales that separate weak from strong coupling of the $3d$ and $5d$ respectively, included in the $7d$ theory. 
 
 \begin{itemize}
 	\item \underline{9d gap equation}
 \end{itemize}
 
The gap equation in 9 dimensions of the Gross-Neveu model at imaginary chemical potential for $b_*=\pi/\beta$, $m_*\not=0$ turns to:

\begin{align}
m_*[-({\cal M}_9\beta^7+D_7(e^{-m_*\beta}))-\frac{\ln^2|z_*|}{4!}(D_5(e^{-m_*\beta})-\frac{4\Gamma_1^5}{525\pi})-\\\nonumber
-\frac{\ln^4|z_*|}{5!}(D_3(e^{-m_*\beta})
+\frac{4\Gamma_2^3}{315\pi})
-\frac{\ln^6|z_*|}{6!}(D_1(e^{-m_*\beta})-\frac{16\Gamma_3}{175\pi})]=0
\end{align}
where $\Gamma=\Lambda\beta$ the cut-off. If we set $\Gamma_1=\Lambda_7\beta$, $\Gamma_2=\Lambda_5\beta$ and $\Gamma_3=\Lambda_3\beta$ and put the $9d$ theory at the critical point ${\cal M}_9=0$ or $G_9=G_{9,*}$ we find from $D_7(e^{-m_*\beta})=0$:

\begin{equation}
\boxed
{m_*\beta=3.68896}
\end{equation}
and that $\Lambda_7\beta=3.8873$ from $D_5(e^{-m_*\beta})-\frac{4(\Lambda_7\beta)^5}{525\pi}=0$, $\Lambda_5\beta=9.8758$ from $D_3(e^{-m_*\beta})+\frac{4(\Lambda_5\beta)^3}{315\pi}=0$ and $\Lambda_3\beta=65.0883$ from $D_1(e^{-m_*\beta})-\frac{16\Lambda_3\beta}{175\pi}=0$.

We will also change (like the $5d$ and $7d$ cases) $\frac{16\Lambda_3\beta}{175\pi}$ with ${\cal M}_3^{(9)}\beta$, $\frac{4(\Lambda_5\beta)^3}{315\pi}$ with ${\cal M}_5^{(9)}\beta^3$ and $\frac{4(\Lambda_7\beta)^5}{525\pi}$ with ${\cal M}_7^{(9)}\beta^5$, which are the mass scales that separate weak from strong coupling of the $3d$, $5d$ and $7d$ respectively, included in the $9d$ theory. All the values of the mass scales were calculated and listed in table 3 as parts of the odd dimensional gap equations we study. For example as we see in table 3, ${\cal M}_3^{(5)}\beta=0.8754$, ${\cal M}_5^{(7)}\beta^3=1.5589$, ${\cal M}_7^{(9)}\beta^5=2.1527$, etc.
 
The higher dimensions gap equations contain the lower dimension equations giving us an idea of what happens to the strong coupling in lower dimensions when in the upper dimension we are at the critical point ${\cal M}_{odd}=0$ (the lower dimensions are weaker coupled but still in  the strong coupling regime, since $G_{odd}>G_{odd,_*})$. The overall picture inside the strong coupling regime for various odd dimensions is:

{\centering
	\begin{tabular}{|c|c|c|c|c|c|c|c|}
		\hline 
		\multicolumn{8}{|c|}{\textbf{Table 2. Inside the strong coupling regime}} \\ 
		\hline 
		Dimensions & $\Lambda_3\beta$ & $\Lambda_5\beta$ & $\Lambda_7\beta$ & $\Lambda_9\beta$ & $\Lambda_{11}\beta$ & $\Lambda_{13}\beta$ & $m_*\beta$ \\ 
		\hline 
		3 & $0$ & $-$ & $-$ & $-$ & $-$ & $-$ & $0.9624=\ln\phi^2$ \\ 
		\hline 
		5 & $4.1252$ & $0$ & $-$ & $-$ & $-$ & $-$ & $2.03185\approx\ln(3\phi^2)$ \\ 
		\hline 
		7 & $16.3642$ & $6.0403$ & $0$ & $-$ & $-$ & $-$ & $2.89218\approx\ln(7\phi^2)$ \\ 
		\hline
		9 & $65.0883$ & $9.8758$ & $3.8873$ & $0$ & $-$ & $-$ & $3.68896\approx\ln(15\phi^2)$ \\ 
		\hline 
		11 & $...$ & $...$ & $...$ & $...$ & $0$ & $-$ & $4.46019\approx\ln(33\phi^2)$\\
		\hline
		13 & $...$ & $...$ & $...$ & $...$ & $...$ & $0$ & $5.21997\approx\ln(71\phi^2)$\\
		\hline
		$\infty$ & $\infty$ &  $....$ & $....$ & $...$ & $...$ & $...$ & $\infty$ \\
		\hline 
	\end{tabular}\par}
	
where $\phi=1.618$ is the golden ratio and the approximation of the $m_*\beta$ condensate mass at $5d$ is $98.5\%$, at $7d$ is $99.4\%$, at $9d$ is $99.5\%$, at $11d$ is $99.9\%$ and at $13d$ is $99.9\%$...

Interestingly, the fermion condensate mass has a general value of the form:

\begin{equation}
\label{mass}
\boxed
{m_*\beta=\ln(\alpha_n\phi^2)}
\end{equation}
where $\alpha_n=1,3,7,15,33,71,151,319...$, ($n=1,2,3..$ and $\alpha_1=1$) \cite{OEIS1}. These values are of the form:

\begin{equation}
\alpha_{n+1}=2\alpha_n+\alpha_k
\end{equation}

where $\alpha_k=1,1,1,3,5,9,17,29...$, ($k=1,2,3..$) from \cite{OEIS} with $\frac{\alpha_{k+1}}{\alpha_{k}}=\phi$ and $\alpha_k$ has the growth rate of the Fibonacci numbers. One may find an interesting approach to a finite temperature fermionic theory where the thermal mass is calculated in \cite{David}. 
We see that as dimension increases the $3d$ theory at its strong coupling version always contributes the $\ln\phi^2$ part. All the other dimensions contribute $\ln3, \ln7, \ln15, ...$. By using the changes in the included $\Lambda$ parts in the $5d$, $7d$ and $9d$ gap equations that mentioned before, with the associated mass scales ${\cal M}$, I end up with the corresponding picture:

{\centering
	\begin{tabular}{|c|c|c|c|c|}
		\hline 
		\multicolumn{5}{|c|}{\textbf{Table 3. Gap equations for arbitrary odd dimensions}} \\ 
		\hline 
		Dimensions & $gap\,\,1$ & $gap\,\,2$ & $gap\,\,3$ & $gap\,\,4$ \\ 
		\hline 
		3 & $D_1(-z*)=0$ & $-$ & $-$ & $-$ \\ 
		\hline 
		5 & $D_1(-z*)-0.8754=0$ & $D_3(-z*)=0$ & $-$ & $-$ \\ 
		\hline 
		7 & $D_1(-z*)-1.3890=0$ & $D_3(-z*)+1.5589=0$ & $D_5(-z*)=0$ & $-$ \\ 
		\hline
		9 & $D_1(-z*)-1.8692=0$ & $D_3(-z*)+3.8933=0$ & $D_5(-z*)-2.1527=0$ & $D_7(-z*)=0$ \\ 
		\hline 
		$\infty$ & $....$ &  $....$ & $....$ & $....$ \\
		\hline 
	\end{tabular}\par}
	
We see that as the dimension increases we need larger values of the cut-off (more energy) to regulate the theories. But according to asymptotic freedom of the Gross-Neveu model, higher energies are equivalent to weaker coupling between the fermions. This does not seem to happen when going from $5$ to $7$ dimensions (at the $9d$ theory). Although the cut-off $\Lambda_5\beta$ is larger than $\Lambda_7\beta$ (so the $5$ dimensional theory is in larger energy conditions and weaker coupling) the gap equation that separates the weak from the strong coupling implies that the $5$ dimensional theory is deeper in the strong coupling regime than the $7$ dimensional theory ($3.8933>2.1527$ or ${\cal M}_5^{(9)}\beta^3>{\cal M}_7^{(9)}\beta^5$). It would be interesting to further examine this anomaly in the future, since it might be another mechanism than strongly coupled dynamics that give rise to Fermi condensation like \cite{Chu} or instanton configurations that behave like particles \cite{Genolini}.	

\section{Free energy at the full bosonisation points $m_*\beta=\ln(\alpha_n\phi^2)$ and $\beta b_*=\pi$}

Let's see for example the free energy density for the model at $5d$ by using a result from \cite{Filothodoros:2018}.

\begin{align}
\label{GNfe5}
\frac{1}{N{\rm Tr}{\mathbb I}_{4}}\Delta f_{(5)} =&-\frac{m_*^2}{2G_{5,*}}+\frac{\beta^4 m_*^4\Gamma}{24\pi^3 \beta^5}+\frac{3}{4\pi^2\beta^5}\left[D_5(-z_*)-\frac{1}{24}\ln^4|z_*|D_1(-z_*)\right]\nonumber\\
&+\frac{\beta b*}{8\pi^2\beta^5}\left[D_4(-z_*)+\frac{1}{6}\ln^2|z_*|D_2(-z_*)\right]\,.
\end{align}

The term that appears in the mass gap equation is $\frac{\beta^4 m^4\Gamma}{24\pi^3 \beta^5}-\frac{1}{32}\ln^4|z_*|D_1(-z_*)$ and it is zero for $m_*=\ln(3\phi^2)$ and $\beta b_*=\pi$ according to the $5d$ case above. For the system to be in the chirally broken phase at $T=0$ I set as usual the corresponding value analogue to ${\cal M}_5$ (instead of the renormalized $G$ part) which is zero at the critical point. The last term is the charge $Q_5$ where at the full bosonisation point is zero and there is an absence of charged excitations with the canonical partition function of the system to be given by the grand canonical partition function of the same system at fixed imaginary chemical potential \cite{Filothodoros:2016txa}. So the free energy at $5d$ and at the full bosonisation point $b_*=\pi/\beta$ and $m_*\beta=\ln(3\phi^2)$ is 

\begin{equation}
\frac{1}{N{\rm Tr}{\mathbb I}_{4}}\Delta f_{(5)}=\frac{3}{4\pi^2\beta^5}D_5(e^{-\ln(3\phi^2)})\rightarrow \frac{\Delta f_{(5)}}{N}=\frac{3}{{\pi^2\beta^5}}D_5(e^{-\ln(3\phi^2)})\approx 3\frac{4\zeta(5)}{5\pi^2\beta^5}
\end{equation}
with $99\%$ accuracy in the result.

I follow a similar procedure for the higher odd dimensions from results in Appendices A and B of \cite{Filothodoros:2018} and I end up with the table above.

{\centering
	\begin{tabular}{|c|c|c|c|c|c|c|}
		\hline 
		\multicolumn{7}{|c|}{\textbf{Table 4. The free energies}} \\ 
		\hline 
		Dimensions & $\frac{\Delta f_{(3)}}{N}$ & $\frac{\Delta f_{(5)}}{N}$ & $\frac{\Delta f_{(7)}}{N}$ & $\frac{\Delta f_{(9)}}{N}$ & $\frac{\Delta f_{(11)}}{N}$ & $m_*\beta$ \\ 
		\hline 
		3 & $\frac{4 \zeta(3)}{5\pi\beta^3}$ & $-$ & $-$ & $-$ & $-$ & $0.9624=\ln\phi^2$ \\ 
		\hline 
		5 & $-$ & $3\frac{4\zeta(5)}{5\pi^2\beta^5}$ & $-$ & $-$ & $-$ & $2.03185\approx\ln(3\phi^2)$ \\ 
		\hline 
		7 & $-$ & $-$ & $15\frac{4\zeta(7)}{5\pi^3\beta^7}$ & $-$ & $-$ & $2.89218\approx\ln(7\phi^2)$ \\ 
		\hline
		9 & $-$ & $-$ & $-$ & $105\frac{4\zeta(9)}{5\pi^4\beta^9}$ & $-$ & $3.68896\approx\ln(15\phi^2)$ \\ 
		\hline 
		11 & $-$ & $-$ & $-$ & $-$ & $945\frac{4\zeta(11)}{5\pi^5\beta^{11}}$ & $4.46019\approx\ln(33\phi^2)$\\
		\hline 
	\end{tabular}\par}
where 

\begin{equation}
\frac{\Delta f_{(d)}}{N}=0.96166/\pi\beta^3, 2.46186/\pi^2\beta^5, 12.13245/\pi^3\beta^7, 85.42275/\pi^4\beta^9, 774.9189/\pi^5\beta^{11}, ...
\end{equation}
for $d=3,5,7,9,11$ dimensions respectively. We observe the fractional number $\tilde{c}=\frac{4}{5}$ that appears in \cite{Sachdev:1993pr} as a factor in all the free energy densities at the full bosonisation point with an approximation with accuracy over $97.6\%$. The generalized form of the free energy densities at the full bosonisation points is:

\begin{equation}
\label{freeenergy}
\frac{\Delta f_{(2n+1)}}{N}=a_n\frac{\tilde{c}\zeta(2n+1)}{\pi^n\beta^{2n+1}}
\end{equation}
where $\zeta(2n+1)$ is the Riemann zeta function, $n=1,2,3,4,..$ and $a_n=1,3,15,105,..,(2n-1)!!$ , with $a_1=1$.

\section{Summary and Discussion}

The main message of this work is that although there are infrared free terms in the gap equations of the odd $d$ Gross-Neveu model at imaginary chemical potential and finite temperature (large $N$ approximation), we can deal with them by using the higher odd dimensional gap equation as a linear combination of the lower dimensional equations. In the whole process I introduce the new idea of the running cut-off $\Lambda_d$ that varies as a function of energy and describes the connection between the theory's parameters with the energy scale. My calculations have unveiled a generalized form of the mass of the fermionic condensate (\ref{mass}) inside the temperature windows of the model and the values of the free energy densities of the form (\ref{freeenergy}) which are analogous to the rational number $\tilde{c}=\frac{4}{5}$ with a very good approximation.

I believe that my results offer a new window into the physics of bosonisation. As I briefly alluded to in the text there is a kind of anomaly in the gap equation in $5d$ which may be due to another mechanism of fermion condensation than the strongly coupled dynamics or instanton configurations. It would be interesting to examine this identity better in future work.

\section*{Acknowledgements}
I would like to thank A. C. Petkou and N. D. Vlachos for their useful comments and help.

\end{document}